\documentclass{moriond}

\usepackage{amsmath}
\usepackage{wrapfig}

\def\be{\begin{equation}}
\def\ee{\end{equation}}
\def\bea{\begin{eqnarray}}
\def\eea{\end{eqnarray}}

\newcommand{\Photo}{\smallskip\includegraphics[height=35mm]{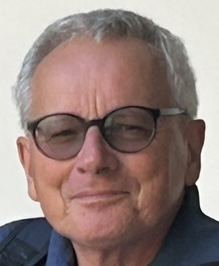}}

\begin{document}
\vspace*{4cm}
\title{Universal properties of elastic $\boldsymbol{pp}$ cross section from the ISR to the LHC}

\author{Micha{\l} Prasza{\l}owicz }

\address{Institute of Theoretical Physics, Jagiellonian University,\\
Łojasiewicza 11, 30-348 Krak{\'o}w, Poland}

\maketitle\abstracts{
We explore the phenomenology of the property that the ratio 
of bump to dip {\em positions} of the elastic differential $pp$ cross section 
is constant over the energy range from the ISR to the LHC. We review 
the old idea of geometric scaling at the ISR and argue that it also holds at the LHC.
We discuss its consequences for the $\rho$ parameter and for the ratio of bump to dip
cross section {\em values}.}

\section{Introduction}

Following Ref.~[1]\nocite{Praszalowicz:2025twy} we argue that the geometrical scaling (GS) discovered at the ISR
still holds at the LHC.
It was shown  Ref.~[2]\nocite{Baldenegro:2024vgg} that ratio of bump to dip {\em postions} 
${\cal T}_{\rm bd}=t_{\rm bump}/t_{\rm dip}$ of the elastic differential $pp$ cross section
is constant over the wide energy range, from 20~GeV (ISR) to 13~TeV (LHC): 
${\cal T}_{\rm bd}=1.355\pm0.011$. This means that dips and bumps have a universal
energy dependence $|t_{\rm dip,\, bump}|=\tau_{\rm dip,\, bump}/R^2(s)$, which cancels
in the ratio. Here $R(s)$ is a so called interaction radius to be determined from the data.
This suggests, that -- up to normalization -- the elastic scattering amplitude $T_{\rm el}(s,t)$,
which formally is a function of two independent variables, $s$ and $t$, is in fact a function
of one dimensionless scaling  variable $\tau=R^2(s)|t|$. This hypothesis is called
geometric scaling.\cite{DiasDeDeus:1973lde,Buras:1973km}

It is instructive to illustrate this in the impact parameter space, where
\begin{equation}
\sigma_{\text{tot}}(s)    =2%
{\displaystyle\int}
d^{2}\boldsymbol{b}\,\operatorname{Im}T_{\text{el}}(s,b), ~~~
\sigma_{\text{el}}(s)    =%
{\displaystyle\int}
d^{2}\boldsymbol{b}\,\left\vert T_{\text{el}}(s,b)\right\vert ^{2}.
\label{eq:sigmas}%
\end{equation}
Inelastic cross section is given as a difference $\sigma_{\text{tot}}(s) -\sigma_{\text{el}}(s)$. 
Note that in this formulation the elastic amplitude is dimensionless. GS in impact parameter
space means that $\operatorname{Im}T_{\text{el}}(s,b)=\operatorname{Im}T_{\text{el}}(B=b/R(s))$.
Changing variables in  (\ref{eq:sigmasb}) 
\begin{equation}
\sigma_{\text{tot}}(s)    =2 R^2(s)%
{\displaystyle\int}
d^{2}\boldsymbol{B}\,\operatorname{Im}T_{\text{el}}(B), ~~~
\sigma_{\text{el}}(s)    = R^2(s)%
{\displaystyle\int}
d^{2}\boldsymbol{B}\,\left\vert T_{\text{el}}(B)\right\vert ^{2}
\label{eq:sigmasb}%
\end{equation}
we conclude that all integrated $pp$ cross sections: elastic, inelastic and total, should have the same
energy behavior, provided one can neglect the real part of the elastic amplitude, which seems
to be justified by the smallness of the $\rho={\rm Re}T_{\rm el}(s,t=0)/{\rm Im}T_{\rm el}(s,t=0)$ 
parameter~\cite{Baldenegro:2024vgg} both
at the ISR and at the LHC. While such uniform energy dependence is in fact seen at the ISR, it is
no longer true at the LHC.
To illustrate this behavior we have fitted~\cite{Baldenegro:2024vgg} power law energy dependence
separately at the ISR and at the LHC
\begin{equation}
\sigma_{\rm tot}^{\rm ISR}\sim \sigma_{\rm el}^{\rm ISR}\sim W^{0.11},~~~
\sigma_{\rm tot}^{\rm LHC}\sim W^{0.17},~~~ \sigma_{\rm el}^{\rm LHC}\sim W^{0.23}\, ,
\label{eq:Wpower}
\end{equation}
where $W=\sqrt{s}$.

Another quantity of interest is ratio 
${\cal R}_{\rm bd}=(d\sigma_{\rm el}/dt|_{\rm bump})/(d\sigma_{\rm el}/dt|_{\rm dip})$
of the cross section {\em values}, which
is strongly energy dependent at the ISR and saturates~\cite{TOTEM:2020zzr} at $\sim 1.8$ at the LHC (see Fig.~\ref{fig:total}).
This means that at the LHC there is a possibility to align dips and bumps by the GS transformation and then
superimpose the values of the cross sections at dips and bumps with a universal transformation $g(s) d\sigma_{\rm el}/dt$. 
Such scaling has been recently proposed~\cite{Baldenegro:2024vgg,Baldenegro:2022xrj}. It turns out that also
at the ISR one can superimpose the  bumps with $g(s)=1/\sigma_{\rm tot}^2(s)\sim1/R^4(s)$, but not dips.\cite{DiasdeDeus:1977af}
 In fact this universality
is violated only in the vicinity of dips. To summarize: GS holds at the ISR with only one universal energy dependent
radius $R^2(s)\sim \sigma_{\rm tot}(s)$, whereas at the LHC the second function $g(s)$ is necessary. In what follows
we show that, nevertheless, GS is present at the LHC, but in a narrower $t$ domain than at the ISR.


\section{Geometric Scaling in momentum space}
	
In momentum space we can rewrite the cross sections (\ref{eq:sigmas}) as
\begin{equation}
s \sigma_{\text{tot}}(s)    =2\operatorname{Im}\tilde{T}_{\text{el}}(s,0),
~~~~
\sigma_{\text{el}}(s)    =\frac{1}{4\pi s^{2}}%
{\displaystyle\int}
dt\left\vert \tilde{T}_{\text{el}}(s,t)\right\vert ^{2} \, ,
\label{eq:sigtotc}%
\end{equation}
where $\tilde{T}_{\rm el}(s,t)$ is a Fourier transform of $T_{\text{el}}(s,b)$ in a normalization where 
 $\tilde{T}_{\rm el}(s,t)$ is dimensionless. Assuming GS and crossing symmetry, the Ansatz for the amplitude
 that reproduces the energy dependence of the total cross section reads~\cite{DiasdeDeus:1975ybq}
\begin{equation}
\tilde{T}_{\text{el}}(s,\tau)=isR^{2}(-is)\Phi\Big( \left\vert t\right\vert
R^{2}(-is)\Big) \, .
\end{equation}
Following the trick of Ref.~[8]\nocite{DiasdeDeus:1975ybq}, which is equivalent to the dispersion relations, we can identify real and imaginary
parts of the amplitude~\cite{Praszalowicz:2025twy}
\begin{equation}
\operatorname{Im}\tilde{T}_{\text{el}}(s,\tau)   = sR^{2}(y)\Phi\left(
\tau\right), ~~~
\operatorname{Re}\tilde{T}_{\text{el}}(s,\tau)   =s\frac{\pi}{2}\frac
{dR^{2}(y)}{dy}\frac{d}{d\tau}\left(  \tau\Phi\left( \tau\right)  \right)  \, ,
\label{eq:ImRe}%
\end{equation}
where $y=\ln s $ is a rapidity. Here $\Phi(\tau)$ is an energy independent, universal function
of the scaling variable $\tau=R^2(s)|t|$.

There are three immediate predictions, which follow from Eqs.~(\ref{eq:ImRe}). The first one
is an absolute prediction for the $\rho$ parameter~\cite{Praszalowicz:2025twy}
\begin{equation}
\rho = \frac{\pi}{2 }\frac{1}{R^{2}(y)}%
\frac{dR^{2}(y)}{dy} \, .
\label{eq:rhopred}%
\end{equation}

The second one is a prediction for ${\cal R}_{\rm bd}$. To this end we have to identify dips and bumps.\cite{Praszalowicz:2025twy}
The dip corresponds to the first zero of $\Phi(\tau_{\rm dip})=0$ and the bump to the first minimum of $\Phi(\tau)$~\footnote{Note that cross sections
are proportional to $\Phi^2$.}. 
At the dip $\operatorname{Im}\tilde{T}_{\text{el}}=0$
and the cross section is entirely given by $\operatorname{Re}\tilde{T}_{\text{el}}$. The bump corresponds
to  $d\Phi(\tau)/d\tau|_{\rm bump}=0$. From these two conditions we obtain~\cite{Praszalowicz:2025twy}
\begin{equation}
{\cal R}_{\rm bd}(s)=c_{0}%
\frac{1+\rho^{2}\left(  y\right)  }{\rho^{2}\left(  y\right)  },
~~~~{\rm where}~~~~
c_{0}=\frac{\Phi^{2}(\tau_{\mathrm{bump}})}{\left(  \tau_{\text{dip}}\frac
{d}{d\tau}\Phi (\tau_{\mathrm{dip}}) \right)  ^{2}} \, .
\label{eq:Rbdc}
\end{equation}

Finally, equations (\ref{eq:ImRe}) allow to predict the energy behavior of the total elastic cross section~\cite{Praszalowicz:2025twy}
\begin{equation}
\sigma_{\text{el}}(s)   
 =\frac{R^{2}(y)}{4\pi}\left(  1+c_{1}\rho^{2}(y)\right)  \times%
{\displaystyle\int}
d\tau{\Phi}^{2}{(\tau)},~~~
{\rm where}~~~
c_{1}=\frac{%
{\displaystyle\int}
d\tau\left(  {\frac{d}{d\tau}\left(  \tau\Phi (\tau)\right)  }\right)
^{2}}{%
{\displaystyle\int}
d\tau{\Phi}^{2}{[\tau]}}.
\label{eq:eltot}
\end{equation}

\section{Phenomenology}

\begin{wrapfigure}{r}{0.48\textwidth}
\vspace{-0.5cm}
\includegraphics[height=5cm]{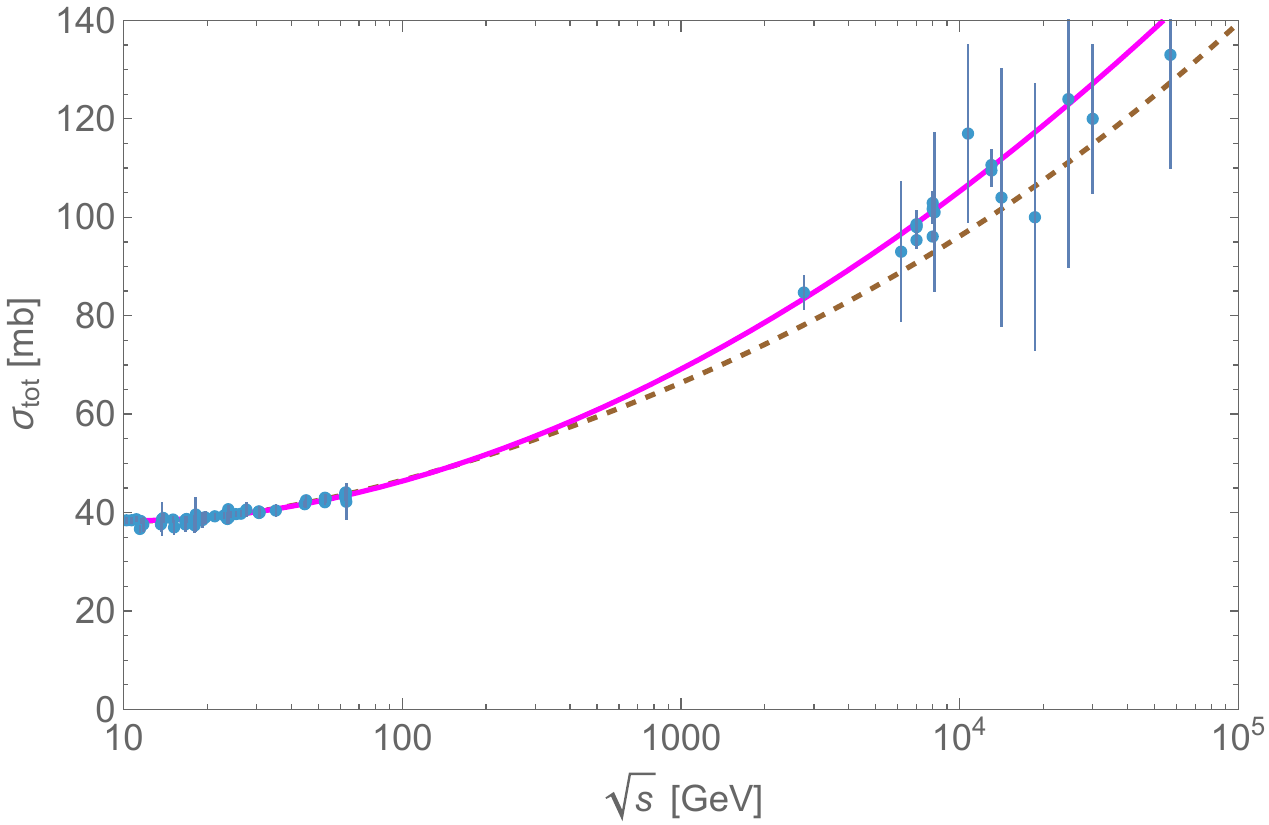}
\caption{Total $pp$
cross-section in mb as a function of $\sqrt{s}$ in GeV. Points at small
$\sqrt{s} < 100$~GeV are from the ISR, points above 1~TeV are from the LHC
(small error bars) and from cosmic rays. Solid magenta line corresponds to the
PDG parametrization,$^{9,10}$
and brown dashed line to Donnachie and 
Landshoff.$^{11}$
} \label{fig:total}
\end{wrapfigure}

Now, we can compare predictions (\ref{eq:rhopred}), (\ref{eq:Rbdc}) and (\ref{eq:eltot}) with data. We choose
$R^2(s)=\sigma_{\rm tot}(s)$. In order to compute $\rho$, ${\cal R}_{\rm bd}$ and $\sigma_{\text{el}}$ we need 
an analytical parametrization of $\sigma_{\rm tot}$.
For illustration purposes we use two analytic parameterizations:\,\cite{Praszalowicz:2025twy}
 the COMPETE parametrization~\cite{Cudell:2001pn} quoted in PDG~(2010)~\cite{PDG2010} and the
older parametrization by Donnachie and Landshoff~\cite{Donnachie:1992ny} devised to describe
lower energy data. They are plotted in Fig.~\ref{fig:total} together with experimental data.\cite{PDG2022}

\begin{figure}[h]
\centering
~\includegraphics[width=7.9cm]{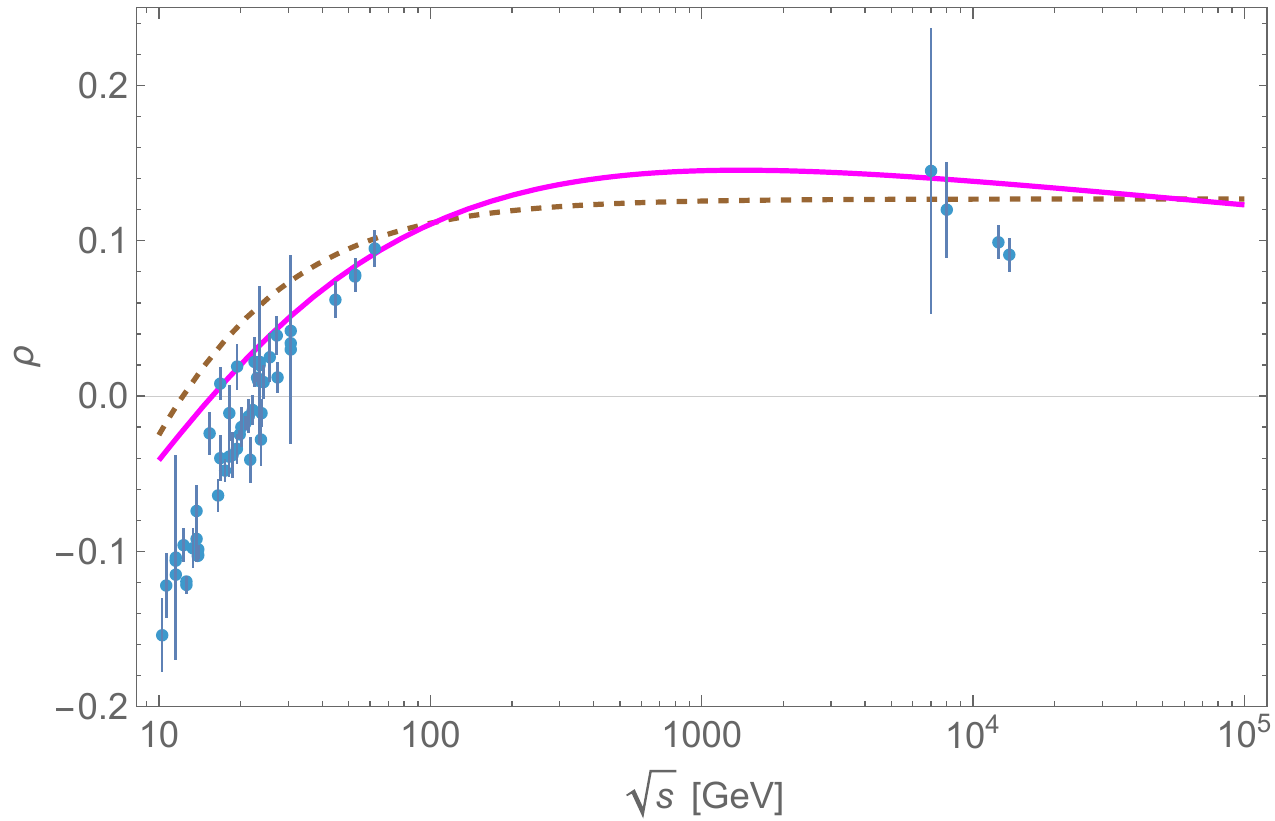}~~~\includegraphics[width=7.7cm]{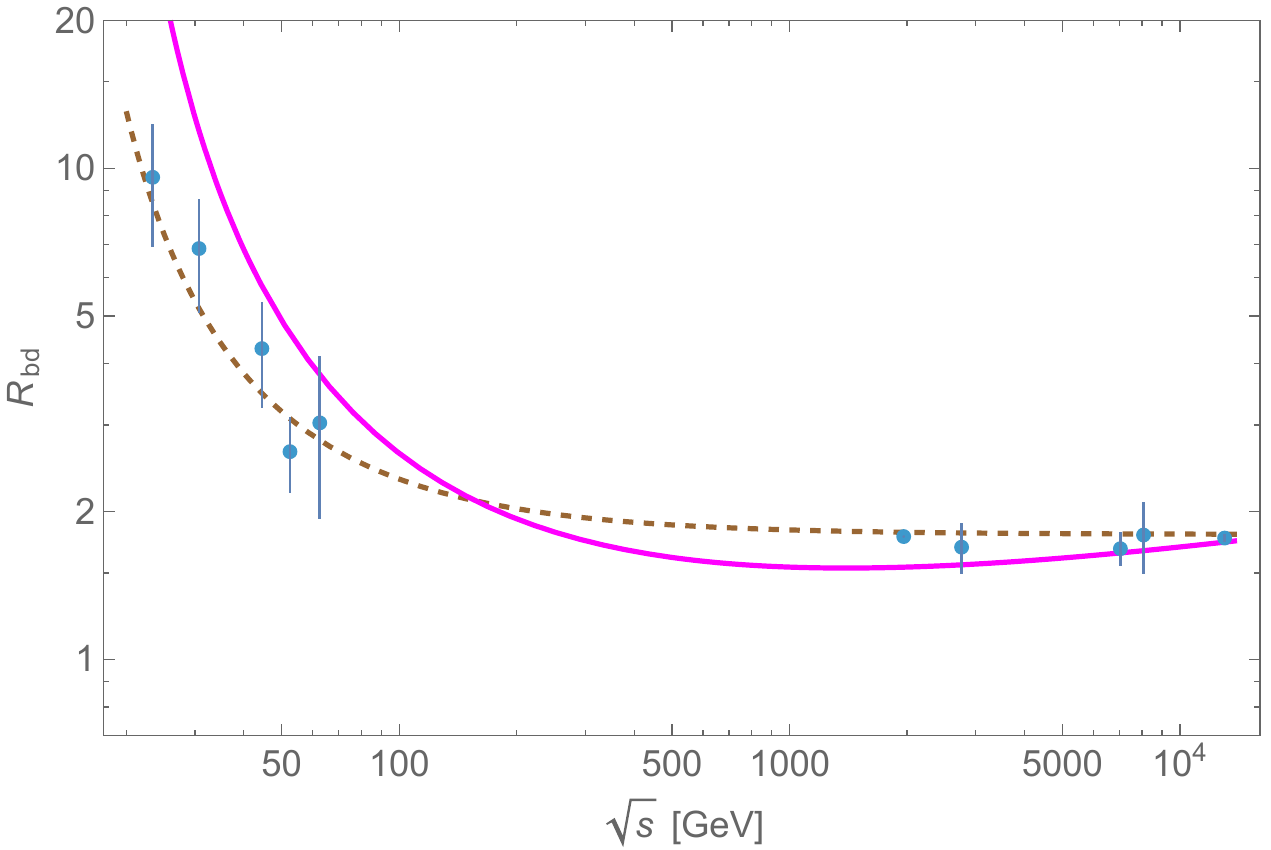} 
\vspace{-0.4cm} 
\caption{Left panel: $\rho$ parameter (\ref{eq:rhopred}). Right panel:  ${\cal R}_{\rm bd}$ ratio (\ref{eq:Rbdc}).}
\label{fig:rhoRbd}%
\end{figure}

In the left panel of Fig.~\ref{fig:rhoRbd} we plot parameter $\rho$ computed according
to (\ref{eq:rhopred}). We see that both low and high energy data
are well reproduced, both normalization and energy dependence. The last two points
in the left panel of   Fig.~\ref{fig:rhoRbd} correspond to
two different estimates of $\rho$ by TOTEM.\cite{TOTEM:2017sdy} Their rapid
decrease with energy  has been attributed the odderon.\cite{TOTEM:2020zzr}

In the right panel of Fig.~\ref{fig:rhoRbd} we plot ${\cal R}_{\rm bd}$ ratio computed according to (\ref{eq:Rbdc}) with one
adjustable parameter $c_0=0.012 \div 0.013$. Data are from Ref.~[5].~\nocite{TOTEM:2020zzr} We see that the DL
parametrization seems to do a better job at the ISR energies, although it is slightly worse that the COMPETE
parametrization for the $\rho$ parameter in this region.

Finally let us comment on the rise of the total elastic cross-section (\ref{eq:eltot}) compared to (\ref{eq:Wpower}).
Note that formula (\ref{eq:eltot}) assumes that GS is valid {\em everywhere} in $t$. This is certainly not true outside of
the dip-bump region, although at the ISR it surprisingly holds even for very small $t$. With this warning in mind
let's observe that
at the ISR the $\rho$ parameter is very small and, despite the fact that it rapidly rises with energy, its influence
on  $\sigma_{\rm el}$ is negligible. At the LHC the $\rho$ parameter is slightly larger but it is almost constant
(see Fig.~\ref{fig:rhoRbd}) and therefore does not modify the energy dependence of  $\sigma_{\rm el}$. We cannot
precisely predict the correction of the energy dependence of $\sigma_{\rm el}$ without knowing the value of 
a constant $c_1$.

\section{Summary}

We have explored the fact that bump to dip positions ratio of the elastic $pp$ cross section is constant
over the wide energy range 20~GeV -- 13~TeV. This suggests the scaling variable $\tau\sim \sigma_{\rm tot} |t|$,
which aligns bump and dip {\em positions} at all energies, known as geometric scaling. By imposing the crossing symmetry
and identifying imaginary and real parts of $\tilde{T}_{\rm el}(s,t)$ we have calculated the $\rho$ parameter
and the ${\cal R}_{\rm bd}$ ratio and compared them with data. We conclude that main properties of total and 
differential cross sections at all energies are explained from a simple and intuitive picture based on GS.
Nevertheless, this picture is only approximate, which is reflected in a failure to reproduce the energy dependence
of $\sigma_{\rm el}$ at the LHC due to the violation of GS at small $t$.


\section*{References}

\end{document}